\documentstyle[aps,twocolumn]{revtex}

\begin{document}
\title{{\bf THE STRONG LEVINSON THEOREM FOR THE DIRAC EQUATION}}
\author{{\bf Alex Calogeracos}$^{*a}$, {\bf Norman Dombey}$^{b}$}
\address{Physics and Astronomy Department,, University of Sussex, Brighton BN1 
9QH,~UK\\
$^{*}$Permanent Address: {\it Division of Theoretical Mechanics, Air Force}\\
Academy TG1010, Dekelia Air Force Base, Greece \\
$^{a}$a.calogeracos@sussex.ac.uk, $^{b}$normand@sussex.ac.uk\\
\flushleft We consider the Dirac equation in one space dimension in the
presence of a symmetric potential well. We connect the scattering phase
shifts at $E=+m$
and $E=-m$ to the number of states that have left the 
positive energy continuum or joined the negative energy continuum 
respectively as the potential is turned on from zero.}
\maketitle

\section{Introduction}

\noindent Levinson \cite{lev} showed in 1949 that the phase shift for
non-relativistic potential scattering in three dimensions at zero energy is
determined by the number of bound states for the particular angular momentum
under consideration. In one dimension a similar result holds for symmetric
potentials although now even and odd parity phase shifts have to be treated
separately. But if a Levinson theorem is considered for the Dirac equation
in either one or three dimensions, authors up to now have only been able to
prove a result in terms of the sum of positive and negative energy phase
shifts at threshold (now defined as zero momentum). For example in one
dimension Lin \cite{lin} considers a symmetric finite range potential $V(x)$
and connects the scattering phase shifts at threshold $E=\pm m$ to the
number of bound states via the relation

\begin{center}
$\Delta _{e,o}(+m)+\Delta _{e,o}(-m)$

$\pm \frac{\pi }{2}\left\{ \sin ^{2}\Delta _{e,o}(+m)-\sin ^{2}\Delta
_{e,o}(-m)\right\} =n_{e,o}\pi$
\end{center}

\begin{equation}  \label{lin1}
\end{equation}

\noindent where the subscripts $e,o$ denote the parity of the state and $%
n_{e,o}$ stands for the number of even and odd bound states. Ma and Ni \cite
{mani} obtain a similar result in three dimensions.

Poliatzky \cite{pol} went further and claimed to have proven a stronger
version of Levinson's theorem for the Dirac equation in three dimensions. He
stated that the phase shift at threshold for a particular $\kappa
,\,j=\left| \kappa \right| -\frac{1}{2}$ (in the usual parametrisation of
the Dirac equation) was given in terms of the number of bound states of a
related Schr\"{o}dinger equation. Ma \cite{ma}, however, pointed out that
this related Schr\"{o}dinger equation did not exist if the potential
strength satisfied $\left| V\right| >2m$. Poliatzky \cite{pol2} accepted
this criticism and blamed pair production for the breakdown of his proof of
a stronger Levinson theorem. He suggested nevertheless that a stronger
version of Levinson's theorem should exist for the Dirac equation even when $%
\left| V\right| >2m.$

In this note we demonstrate the existence of a stronger version of
Levinson's theorem for the Dirac equation. For simplicity we choose to do
this in one dimension and follow Lin by considering finite range potential
wells $V(x)$ which are symmetric and integrable at the origin and where the
potential is the zeroth component of a four vector. We obtain {\it separate }%
results for the phase shift at $E=+m$ and at $E=-m$. In particular our
results connect the phase shift at $E=+m$ to the number of states that have
left the positive energy continuum to become bound by the potential, while
the phase shift at $E=-m$ is connected to number of states that have joined
the negative energy continuum. These results constitute the stronger version
of the theorem which we shall just call the strong Levinson theorem as
opposed to the weak version above. Our approach uses Lin's results \cite{lin}
for the behaviour of $\tan \Delta $ at small momentum $k$ together with a
simple generalisation of the counting argument used by Barton \cite{barton}.
At no point do we need to consider any related Schr\"{o}dinger equation nor
are there any limitations on the strength of the potential.

We consider an electron in a box extending from $x=-L$ to $x=L$ and
eventually take the limit $L\rightarrow \infty $. In this limit states that
correspond (for $L$ finite) to a real wavevector $k$ with energy $\left|
E\right| >m$ become continuum scattering states, whereas states that
correspond to imaginary wavevectors become bound states. This technique of
box normalization in the case of the Dirac equation has previously been
employed by Bell and Rajamaran \cite{bell}; more recently it was used in the
context of the Schr\"{o}dinger equation \cite{mats} to connect boundary
conditions with phase shifts just as we will do here.

For definiteness we consider a potential $V_{0}(x)$ which satisfies $%
V_{0}(x)\leq 0$ everywhere with $V_{0}(x)=0,\left| x\right| >a$ for some $%
a<L.$ We follow Lin and assume that the potential is switched on
adiabatically starting with $V=0$ by considering the potential $\lambda
V_{0}(x)$ and letting $\lambda $ vary smoothly from $\lambda =0$ to $\lambda
=1$. The advantage of putting the system in a box and of the consequent
discrete spectrum is that we can follow the energy eigenvalue of each state
of a given $n$ as the potential deepens. This will allow us to show the
behaviour of the even or odd phase shift $\Delta _{e,o}(+m)$ as a state
crosses the value $E=+m$ and will give the strong Levinson theorem for
positive energies. In contrast, however, to the Schr\"{o}dinger case,
knowledge of $\Delta _{e,o}(+m)$ is not sufficient to determine the number
of bound states. As the potential deepens further a bound electron state can
cross the value $E=0$ and it eventually reaches $E=-m$ where it joins the
negative energy continuum, since in one dimension the states at $\left|
E\right| =m$ are not normalizable.

In addition to the scattering states $\left| E\right| >m$ and bound states $%
\left| E\right| <m$ for completeness we have to consider the special case of
critical states that occur for certain values of the potential: these are
the states called zero energy resonances or half-bound states \cite{Newt} in
the context of the Schr\"{o}dinger equation, but in the Dirac equation these
include states at both $E=-m$ as well as $E=m$ \cite{CDK},\cite{kd}. We deal
with critical potentials in the final section of the paper.

\section{The strong Levinson theorem in the absence of critical states}

We write the Dirac equation as a two-component spinor $u=\left( 
\begin{array}{l}
f \\ 
g
\end{array}
\right)$ where \cite{CDK}

\begin{mathletters}
\begin{eqnarray}
\frac{\partial f}{\partial x}+(E-V(x)+m)g &=&0 \\
\frac{\partial g}{\partial x}-(E-V(x)-m)f &=&0
\end{eqnarray}

\noindent and then the even and odd free particle wavefunctions outside the
well $x>a$ for positive energies are

\end{mathletters}
\begin{mathletters}
\label{whole}
\begin{eqnarray}
u_{e}(k,x) &=&\frac{{\cal N}_{e+}(k)}{\sqrt{L}}\left( 
\begin{array}{c}
\cos \left( kx+\Delta _{e}\right) \\ 
\frac{k}{E+m}\sin \left( kx+\Delta _{e}\right)
\end{array}
\right)  \label{6} \\
u_{o}(k,x) &=&\frac{{\cal N}_{o+}(k)}{\sqrt{L}}\left( 
\begin{array}{c}
\sin \left( kx+\Delta _{o}\right) \\ 
-\frac{k}{E+m}\cos \left( kx+\Delta _{o}\right)
\end{array}
\right)
\end{eqnarray}
with corresponding expressions for $x<a$ where $e,o$ refer to even and odd
states. The bound state wavefunction for $x>a$ is given by 
\end{mathletters}
\begin{equation}
u_{b}(x)={\cal N}_{b}e^{-\kappa x},x>a  \label{8}
\end{equation}

\noindent where $\kappa =\sqrt{m^{2}-E_{b}^{2}}$ and for $x<-a$ according to
its parity. The ${\cal N}$ are normalisation factors and play no role in the
discussion.

We now use Lin's results for the behaviour of the phase shifts at threshold $%
k\rightarrow 0$ in the continuum case (i.e. without box normalisation) and
in the {\it absence }of critical states. For a wavevector $k$ we define $\xi
\equiv ka$. In the limit $\xi \rightarrow 0$ Lin finds that for positive
energies 
\begin{mathletters}
\label{d}
\begin{eqnarray}
\tan \Delta _{e}(+m) &\sim &\xi ^{-\left( 2\widetilde{p}_{e}-1\right) } \\
\tan \Delta _{o}(+m) &\sim &\xi ^{2\widetilde{p}_{o}-1}
\end{eqnarray}
and for negative energies

\end{mathletters}
\begin{mathletters}
\label{even}
\begin{eqnarray}
\tan \Delta _{e}(-m) &\sim &\xi ^{2\widetilde{q}_{e}-1} \\
\tan \Delta _{o}(-m) &\sim &\xi ^{-\left( 2\widetilde{q}_{o}-1\right) }
\end{eqnarray}
where $\widetilde{p}_{e,o},\widetilde{q}_{e,o}$are positive integers. This
will allow us to determine whether a phase shift at threshold is $0$ $%
\mathop{\rm mod}%
\pi $ or $\pi /2$ $%
\mathop{\rm mod}%
\pi $ and in particular shows that

\end{mathletters}
\begin{equation}
\sin ^{2}\Delta _{e,o}(+m)-\sin ^{2}\Delta _{e,o}(-m)=\pm 1
\end{equation}
so the weak Levinson theorem in the absence of critical states is from Eq. (%
\ref{lin1})

\begin{equation}
\Delta _{e,o}(+m)+\Delta _{e,o}(-m)+\pi /2=n_{e,o}\pi  \label{weak}
\end{equation}

We now turn to the counting argument. We impose the periodic boundary
condition 
\begin{equation}
\psi (-L)=\psi (L).  \label{4aa}
\end{equation}
Since the potential is symmetric, parity conservation requires that the top
component of an even spinor is even as is the bottom component of an odd
spinor. Eq. (\ref{4aa}) thus makes a non-trivial statement only about odd
components and gives for both even and odd states 
\begin{equation}
\sin \left( kL+\Delta _{e,o}\right) =\sin \left( -kL-\Delta _{e,o}\right) =0
\label{16a}
\end{equation}

\noindent and hence 
\begin{equation}
k_{n}L+\Delta _{e,o}(k_{n})=n\pi  \label{16}
\end{equation}

\noindent where $n$ is an integer.

If we repeat this analysis for a free particle where $V(x)=0$ for all $x$ we
obtain 
\begin{equation}
k_{n}L=n\pi
\end{equation}

\noindent as expected. Since $k$ is nonnegative by definition, so is $n$
while Eqs. (\ref{whole}) shows that $k=n=0$ is allowed. Thus for a free
particle in a box states are labelled by integer values $n=0,1,2,..$ with
corresponding values for the wavevector $k_{n}=0,\pi /L,2\pi /L,..$

Assume a potential well of the form $V(x)=\lambda V_{0}(x)$, where we take $%
V_{0}(x)\leq 0$ in order to ensure that there is at least one bound state. $%
V_{0}(x)$ is symmetric and is less singular than $1/x$ at the origin. $%
\lambda $ starts at $\lambda =0$ with $\Delta _{e,o}=0$ and increases to $%
\lambda =1$. Lin \cite{lin} shows that the phase shift dependence on $%
\lambda $ is given by

\begin{equation}
\frac{d\Delta _{e,o}(E)}{d\lambda }=-\frac{\pi E}{k}(u_{e,o}(\lambda
),V_{0}u_{e,o}(\lambda ))\geq 0
\end{equation}
where the outer brackets denote the inner product. So near threshold$%
\,\Delta _{e,o}(m+k^{2}/2m)\geq 0.$ Let us begin with positive energy even
states. From Eq.(5a), $\tan \Delta _{e}(\xi )\rightarrow \infty $ as $\xi
\rightarrow 0$ and so $\Delta _{e}(+m)$ must be a positive odd multiple of $%
\pi /2$ or 
\begin{equation}
\Delta _{e}(+m)=\left( \mu _{+}-\frac{1}{2}\right) \pi  \label{strong}
\end{equation}
where $\mu _{+}$ is a positive integer (the subscript refers to the sign of
energy) whose physical meaning we find shortly. Substituting Eq.(\ref{strong}%
) in Eq. (\ref{16}) gives for values of $k$ near threshold 
\begin{equation}
k_{n}=\frac{\pi }{L}\left( n-\mu _{+}+\frac{1}{2}\right)  \label{5b}
\end{equation}

\noindent Since $k_{n}$ is by definition non-negative 
\begin{equation}
n\geq \mu _{+}  \label{cond}
\end{equation}

\noindent Note that when $V(x)=0,$ $n\geq 0$. To determine the meaning of $%
\mu _{+}$ compare $\lambda =0$ to a potential well with a very small value $%
\lambda =\varepsilon >0$. There is a one-to-one correspondence between
states for different values of $\lambda $ and the energy levels $E_{n}$ for $%
n\geq 1$ are hardly changed by the potential. But the energy level $E_{n}$
for $n=0$ becomes a bound state with energy $E_{0}<m$ when $\lambda
=\varepsilon $. So for this small value of $\lambda ,\,\mu _{+}=1$ and one
state has crossed $E=m$ from the continuum and become bound. Therefore from
Eq. (\ref{strong}) the strong Levinson theorem is given by

\begin{equation}
\Delta _{e}(+m)=\pi /2
\end{equation}
in agreement with the Schr\"{o}dinger result for a weak potential well. Now
increase $\lambda $ gradually. For some value of $\lambda $ another state
becomes almost bound. We deal with critical states in the next section so
increase $\lambda $ sufficiently in order that the new potential possesses $2$
bound states. Now the continuum states would satisfy $n\geq 2$ and hence $%
\mu _{+}=2.$ So the strong Levinson therem for this potential is $\Delta
_{e}(+m)=3\pi /2$. Thus we now can see that Eq. (\ref{strong}) gives the
strong Levinson theorem where $\mu _{+}$ even states have crossed from the
positive energy continuum and become bound as $\lambda $ has increased from $%
0$ to $1$.

Now consider odd states. A weak potential does not bind odd states and
consequently $\Delta _{o}(+m)=0$ for $\lambda =\varepsilon .$ From Eq. ($5b)$
we see that in general

\begin{equation}
\Delta _{o}(+m)=\nu _{+}\pi  \label{22}
\end{equation}

\noindent where $\nu _{+}$ is a non-negative integer. Then since from Eq. (%
\ref{16}) $k_{n}=\frac{\pi }{L}\left( n-\nu _{+}\right)$ we now have $n \geq
\nu _{+}$ which implies that $\nu _{+}$ states have crossed $E=m$ and become
bound as $\lambda $ increases from $0$ to $1$. Eq. (\ref{22}) is the strong
Levinson theorem for odd states. Similar expressions for the strong form of
Levinson's theorem for even and odd states at positive energies have been
obtained by Farhi, Graham, Jaffe and Weigel \cite{jaffe} in the context of
field theory.

If $\mu _{+},\nu _{+}$ were the number of even and odd bound states of the
potential these results would agree with Levinson's theorem for the
Schr\"{o}dinger equation \cite{sas}. For strong potentials, however, this is
not the case in the Dirac equation because some of these states will
subsequently become supercritical as $\lambda $ increases, cross $E=-m$ and
revert to the continuum. To illustrate this, consider the phase shifts for
negative energy states $\Delta _{e,o}(-m).$ From the weak theorem Eq. (\ref
{weak}) we have

\begin{equation}
\Delta _{e}(-m)=n_{e}\pi -\pi /2-\left( \mu _{+}-\frac{1}{2}\right) \pi
=(n_{e}-\mu _{+})\pi
\end{equation}
The difference $\mu _{+}-$ $n_{e}$ between the number of states $\mu _{+}$
which have crossed $E=m$ as $\lambda $ has increased from $0$ to $1$ and the
number of bound states $n_{e}$ for a given potential with $\lambda =1$ is
just the number of states $\mu _{-}$ which have become supercritical and
crossed $E=-\dot{m}$. So the strong Levinson theorem for negative energy
even states is

\begin{equation}
\Delta _{e}(-m)=-\mu _{-}\pi  \label{negeven}
\end{equation}
and for negative energy odd states

\begin{equation}
\Delta _{o}(-m)=-(\nu _{-}+\frac{1}{2})\pi   \label{negodd}
\end{equation}
where $\nu _{+}-n_{o}=\nu _{-}$ and $\nu _{-}$ is the number of odd states
which have crossed $E=-m$ into the negative energy continuum.

It is perhaps worth commenting that we are using Lin's definition \cite{lin} 
of the phase shift 
which differs from that normally used in non-relativistic
scattering where the phase shift at finite energy $E$ is compared with that at
$E=\infty$ where the phase shift is taken to be zero. This is done in order to 
remove ambiguities of multiples of $\pi$ in the value of the phase shift. But
Parzen showed over 50 years ago \cite{parz} 
that the phase shift at infinite energy in the Dirac equation in three dimensions
for any fixed angular momentum $j$ is given by

\begin{equation}
\delta_{j} (\infty )=-\int_{0}^{\infty }V(r)dr
\label{phase}
\end{equation}
and so phase shifts cannot vanish at infinity in the Dirac equation. Lin
defines his zero of phase shift to be at $V=0$ and then turns on the
potential. In one 
dimension he obtains the analogous relation to Eq. (\ref{phase})

\begin{equation}
\delta (\infty )=-\int_{-\infty }^{\infty }V(x)dx
\label{1D}
\end{equation}
Lin shows in detail that his method also removes any ambiguity in the 
definition of the phase shift since Eq. (\ref{1D}) needs to be satisfied.

\section{The effect of critical states}

For completeness we now must consider critical states. We need the analogues
of Eqs. (\ref{d}, \ref{even}) for the phase shifts of a critical state for $%
\xi \rightarrow 0$ For positive energy critical states Lin \cite{lin} shows

\begin{mathletters}
\label{critpos}
\begin{eqnarray}
\tan \Delta _{e}(+m) &\sim &\xi ^{2p_{e}-1}  \label{crit} \\
\tan \Delta _{o}(+m) &\sim &\xi ^{-\left( 2p_{o}-1\right) }
\end{eqnarray}
and for negative (even or odd) energy states

\end{mathletters}
\begin{mathletters}
\label{super2}
\begin{eqnarray}
\tan \Delta _{e}(-m) &\sim &\xi ^{-\left( 2q_{e}-1\right) }  \label{critneg}
\\
\tan \Delta _{o}(-m) &\sim &\xi ^{2q_{o}-1}
\end{eqnarray}

\noindent where the quantities $p_{e,o}$, $q_{e,o}$ are positive integers.

Consider $\Delta _{e+}(m)$. According to Eq.(\ref{strong}) $\Delta
_{e+}(m)=\pi /2$ for small $\lambda $ and passes through the values $3\pi /2$%
, $5\pi /2$ as $\lambda $ increases and states cross the $E=+m$ level and
become bound. During the process $\Delta _{e+}(m)$ goes through the values $%
\pi $, $2\pi $,.. . These are precisely the values at which (according to
Eq.(\ref{crit})) the potential supports an even positive energy critical
state. Thus the strong version of Levinson's theorem reads as follows: if $%
\mu _{+}$ is the number of positive energy even states that have crossed $%
E=+m$ and the next even state is critical with $E=+m$ then 
\end{mathletters}
\begin{equation}
\Delta _{e}(+m)=\mu _{+}\pi  \label{10}
\end{equation}
Similarly for odd critical states. The phase shift $\Delta _{o+}(m)$ starts
at zero and goes through the values $\pi $, $2\pi $,$..$ as the potential
deepens and positive energy odd states cross $E=+m$ and become bound. Thus $%
\Delta _{o+}(m)$ goes through the values $\pi /2$, $3\pi /2$,$...$ and
according to Eqs. (\ref{critpos}) these are the values where the potential
supports an odd critical state at $E=+m$. Thus the strong Levinson theorem
states that if $\nu _{+}$ is the number of positive energy odd states that
have crossed $E=+m$ and the next odd state in line sits at $E=+m$ then 
\begin{equation}
\Delta _{o}(+m)=\left( \nu _{+}+\frac{1}{2}\right) \pi  \label{23}
\end{equation}
These results agree with those of Sassoli di Bianchi \cite{sas} for the
Schr\"{o}dinger equation with the replacement of $\mu _{+},\nu _{+}$ by $%
n_{e},n_{o}.$

For negative energies we find that if $\mu _{-}$ is the number of even
states that have crossed $E=-m$ and there is a new even supercritical state
at $E=-m$ then 
\begin{equation}
\Delta _{e}(-m)=-\left( \mu _{-}+\frac{1}{2}\right) \pi  \label{18}
\end{equation}
and that if $\nu _{-}$ is the number of odd states that have crossed $E=-m$
and become supercritical and there is a new odd supercritical state at $E=-m$
then 
\begin{equation}
\Delta _{o}(-m)=-\left( \nu _{-}+1\right) \pi  \label{27}
\end{equation}

\section{Conclusions and Acknowlegements}

We have shown that there is a strong Levinson theorem for  
the Dirac equation which allows the
scattering phase shift at threshold to be determined for both positive and
negative energies and both even and odd parities. The phase shift at
threshold, however, does not determine the total number of bound states when
the potential is strong. This is on account of what mathematicians call
spectral flow: a strong attractive potential can become supercritical and
thus states which had been bound can revert to the (negative energy)
continuum.

We expect that our results can be extended to more general potentials
without too much trouble. Nevertheless it seemed important to us to
demonstrate for a simple class of potentials that the strong Levinson's
theorem does exist in the context of the Dirac equation given that previous
attempts going back to the 1960s have been unsuccessful \cite{bart}, \cite
{clem}, \cite{pol} and at best have shown the weak theorem. In addition we
should emphasize that our results explain why a strong Levinson theorem for
the Dirac equation in terms of the number of bound states is not always
possible for potentials where $\left| V\right| >2m$.

We thank G. Barton, D. Waxman and R. L. Jaffe for discussions. One of us
(AC) wishes to thank the Physics and Astronomy Department of the University
of Sussex for its hospitality.
The other (N.D) thanks the Santa Fe Institute
for its hospitality, where some of this work was performed.

\noindent

\end{document}